\documentclass[twocolumn,amsmath,amssymb]{revtex4}

\usepackage{graphicx}
\usepackage{color}
\usepackage{mathtools}
\begin{document}
\title{Crumple-Origami Transition for Twisting Cylindrical Shells}
\author{Li-Min Wang$^1$, Sun-Ting Tsai$^2$, Chih-yu Lee$^3$, Pai-Yi Hsiao$^4$, Jia-Wei Deng$^1$, Hung-Chieh Fan Chiang$^1$, Yicheng Fei$^5$,  and Tzay-Ming Hong$^{1\dagger} $}
\affiliation{$^1$Department of Physics, National Tsing Hua University, Hsinchu 30013, Taiwan, Republic of China}
\affiliation{$^2$Department of Physics and Institute for Physical Science and Technology, University of Maryland, College Park, MD 20742}
\affiliation{$^3$Hsinchu Senior High School, Hsinchu 30013, Taiwan, Republic of China}
\affiliation{$^4$Department of Engineering and System Science, National Tsing Hua University, Hsinchu 30013, Taiwan, Republic of China}
\affiliation{$^5$Department of Physics and Astronomy, Rice University, Houston, TX 77005}

\date{\today}
\begin{abstract}
Origami and crumpling are two extreme tools to shrink a 3-D shell. In the shrink/expand process, the former is reversible due to its topological mechanism, while the latter is irreversible because of its random-generated creases. We observe a morphological transition between origami and crumple states in a twisted cylindrical shell.  By studying the regularity of crease pattern, acoustic emission and energetics from  experiments and simulations, we develop a model to explain this transition from frustration of geometry that causes breaking of rotational symmetry. In contrast to solving  von K$\rm\acute{a}$rm$\rm\acute{a}$n-Donnell equations numerically, our model allows derivations of analytic formula that successfully describe the origami state. When generalized to truncated cones and polygonal cylinders, we explain why multiple and/or reversed crumple-origami transitions can occur. 
\end{abstract}
\maketitle
In contrast to crumpling, origami\cite{Origamiweb} produces regular creases on a membrane. By use of predefined ordered patterns or spontaneous buckling\cite{Kresling}, it can be exploited to design deployable and robust deformable structures\cite{Miura fold, li battery, muscle}, self-folding machines\cite{selffolding}, tunable metamaterials\cite{metamaterials}, DNA origami\cite{DNA}, curved structures\cite{curved origami},  etc. From the geometrical perspective, these structures have the hallmark of topologically protected behavior\cite{topology} and excellent capability of shrinking volume. 

Unlike origami, crumpling generates random creases and configurations\cite{Deterministic Folding, ordered phase} that can not be predicted by classical mechanics. However, it still follows several statistical rules, such as power-law behavior in acoustic emission\cite{crumpling-acoustic, AIC} and force response\cite{crumpling-compression, crumpling-comp and geo}, log-normal distribution for crease length\cite{geo of crumpling}, and logarithmic decay of size and height of crumpled ball with time under compression\cite{crumpling-comp and geo, crumpling-time}. In the initial stage of crumpling, the kite model by Witten\cite{single-ridge} can predict the ratio of stretching and bending energies on each ridge and how their sum increases with the ridge length. As ridge-ridge interactions become important, how the resistence force, average length and number of ridges vary with the radius of crumpled ball have also been deduced by theory\cite{compact crumple}, molecular dynamics simulation and experiments\cite{ordered phase, compact crumple, ridge-ridge}.

Although origami and crumple exhibit distinct characteristics, we are interested in knowing whether these  two morphological states can coexist in the same physical system under certain form of external stress. Related researches are the determination of symmetric/non-symmetric folds in crumpling\cite{Deterministic Folding}, wrinkle-to-crumple transition of elastic sheets on a liquid drop\cite{wrinkle-crumple}, wrinkle-to-fold transition of elastic interface under lateral compression\cite{wrinkle-fold}, and creasing\cite{creasing}. Because origami can be considered as a regular arrangement of creases, it seems likely that, if we can combine wrinkle-to-fold and wrinkle-to-crumple with creases developed from symmetric/non-symmetric folds, there may exist a transition between origami and crumple states. Ostensibly this attempt is hard because these two transitions are defined by different order parameters and belong to distinct physical systems. To describe the crease pattern-formation of deformed membranes, past theories have to rely on numerical or asymptotic method to solve von K$\rm\acute{a}$rm$\rm\acute{a}$n, von K$\rm\acute{a}$rm$\rm\acute{a}$n-Donnell, and Koiter equations\cite{crumpling-theory, Landau, Donnell, Curvature-induced, local buckling} in flat sheets, cylindrical and curved shells, respectively. It is challenging to find exact solutions by these two approaches even for a flat membrane, much less solve cases with arbitrary shapes. Despite of these obstacles, references \cite{wrinkle-crumple, wrinkle-fold} provide studying localization of stress as a clue to incorporate the transition between origami and crumple states. In the following, we will present the finding of a physical system where these two states coexist, and provide a phenomenological theory that not only enables exact solutions, but can be generalized to tackle different shapes of shell without invoking complicated differential geometry.

\begin{figure}[!h]
	\centering
	\includegraphics[width=8cm]{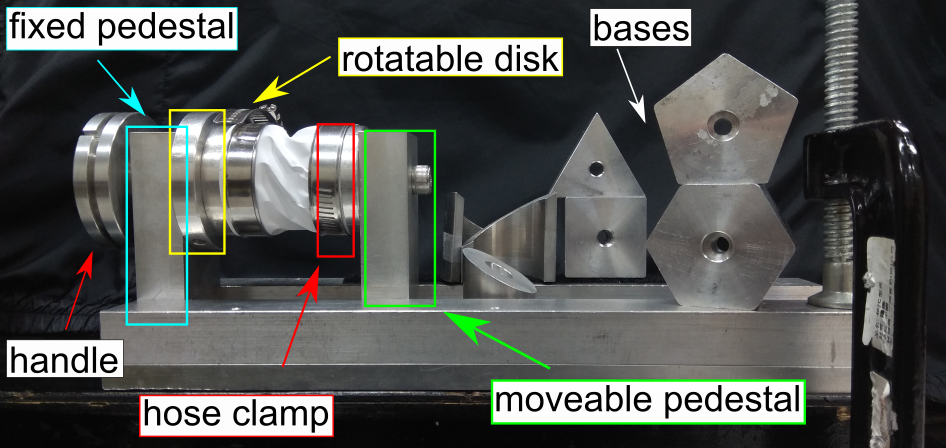}
	\caption
	{(color online) The base of twist machine can be switched from circular and polygonal. Hose clamps are employed to attach and seal the paper shell onto the base.  Extra circular segments are inserted between the clamps and sample to get a better grip on the polygonal base.
	}\label{machine} 
\end{figure}

{\sl Experiment.---} The setup of twisting machine is shown in Fig. \ref{machine}. The shape of its base can be switched from circular to polygonal. Various lengths $L$ are tested for each cylindrical shell, and the sample material ranges from paper to plastic to metal foils of copper or aluminum. The pedestals are kept low to avoid friction between the movable pedestal and track due to torque generated by the shrinkage of twisted shell. 

\begin{figure}[!h]
	\centering
	\includegraphics[width=8cm]{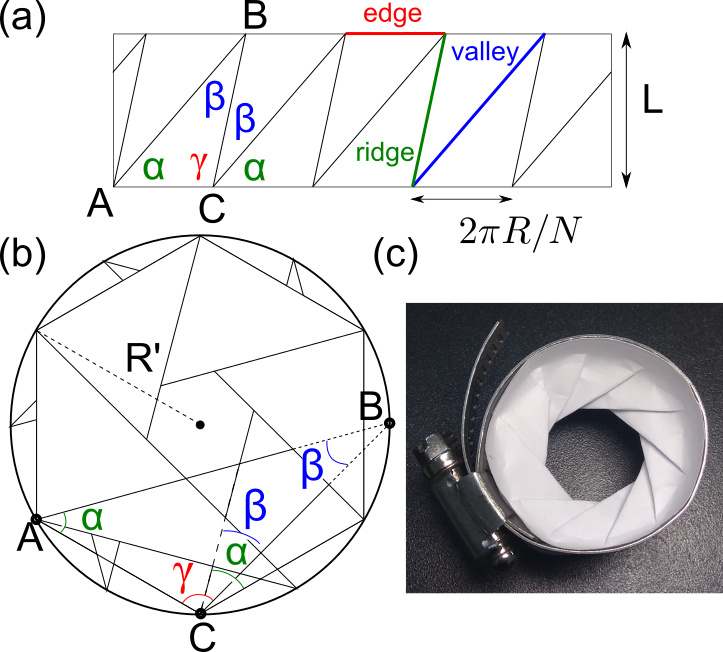}
	\caption
	{(color online) Schematic plots in (a, b) shows the geometry of creases after and bepfore a twisted cylindrical is unfolded. Panel (c) shows the real sample.
	}\label{alpha_theory} 
\end{figure}

In the first phase of this project we appealed to hand twisting for quick results. According to material science, deformation will start from the boundaries of shell where external torque is applied and develop into buckling. We were puzzled when the creases that appeared upon more twisting did not follow a regular pattern nor as reported by Ref. \cite{Kresling}. Intuitively the first culprit for the irregularities is human errors, such as unsteadiness of hands or misalignment of the twisted cylinder. This prompted the design and use of twisting machine. However, the irregularities remain. After more observations we conclude that the regular structure only applies to short cylinders. 

When $L$ is shorter than the diameter $2R$, $N$-pairs of valleys and ridges\cite{Kresling} appear almost simultaneously, indistinguishable by high-speed camera with 6000 fps, as shown schematically in Fig. \ref{alpha_theory} (a). In the mean time, a third kind of crease, which we termed edges, developed on the boundaries of cylindrical shell as they are deformed into regular $N$-polygons. 

\begin{figure}[!h]
	\centering
	\includegraphics[width=8cm]{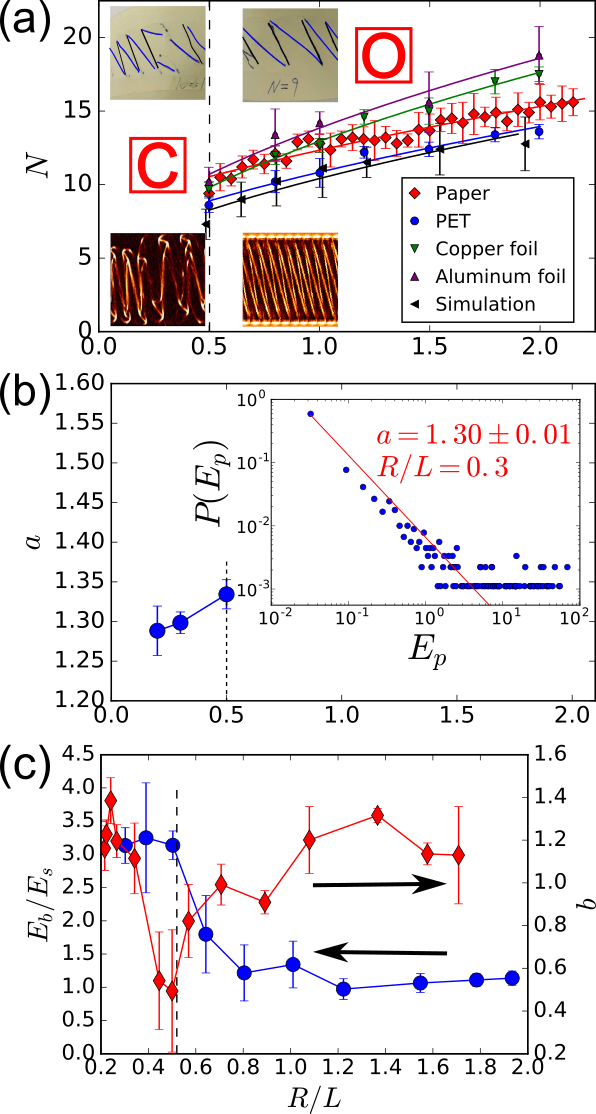}
	\caption
	{(color online) Evidence for the transition of C/O state, marked by red squres, at $R/L=0.5$ with external tension 302 gram weights.  Panel (a) demonstrates that $N$ can be well fit by  $\sqrt{c(R/L)+d}$ where $c, d$ vary with material. Panel (b) shows how the exponent of  $P(E_p)\sim 1/(E_p)^a$ varies with $R/L$. The inset demonstratespower-law behavior where error bars $\delta a$ are the standard error of maximum likelihood estimator. Circles and diamonds  in (c)  depict the $R/L$-dependence of  $E_b/E_s$ and $b$ from MD simulation. 
	}\label{material} 
\end{figure}

Unlike short cylinders, deformations for $L>2R$ progressed from the boundaries are not symmetric and caused breakout of local buckling that is random in both time and space. The valleys and ridges that ensue not only differ in length, but also may segment into multiple pieces, as shown in Fig. \ref{material} (a).

Figure \ref{material} (a) shows that   $N=\sqrt{c(R/L)+d}$ only depends on the ratio\cite{exp data} of $R$ and $L$, and is not sensitive to the twisting speed  $ 0.05\sim 0.5$ revolutions per second. For short cylinders, ridges/valleys and edges compose a periodic triangular pattern that makes the shell deployable. We denote this as an origami state to distinguish from the crumple state that describes the irregular and unfoldable pattern on long cylinders. 

This crumple-origami (CO) transition at $R/L\approx 0.5$ leaves its signature at acoustic emission. Our recording setup and detailed algorithms for counting the pulses have been described in Ref. \cite{chopsticks}. Different from Ref. \cite{crumpling-acoustic}, the cylinder was twisted in only one direction. More than 40 pulses can be gathered for each trial in the crumple state, compared to roughly 2 in the origami state. The inset of Fig. \ref{material} (b) demonstrates that probability density function $P(E_p)$ follows the power law: $P(E_p)\sim 1/(E_p)^{a}$ where $E_p$ denotes the pulse energy - reminiscent of the crumpling process\cite{crumpling-acoustic, AIC}. The exponent $a=1.30 \sim 1.34$ in the range of $R/L=0.3 \sim 0.5$ is determined by the maximum likelihood estimation, similar to $a$ when crumpling the same material. 

The simultaneous formation of ridges/valleys limits the number of sounds short cylinders can emit. Once the origami state is formed,  the topological mechanism\cite{topology} protects the faces between creases from further buckling and thus no more emission of crackling noise. In contrast, the twisting of long cylinders share the same local buckling as crumpling processes, which occurs randomly in place and time. Therefore, it comes as no surprise that they  share the same statistics for acoustic emission. A toy model of drums was offered in Ref. \cite{AIC} to explain the origin of this power law.

{\sl Theoretical model.---} Why is the pattern of creases in the origami state triangular? We can argue it from three experimental facts in this state. (1) Twisting symmetry: twisting one end by $\theta$ is indistinguishable from twisting both ends by $\pm{\theta/2}$, (2) creases will follow the shortest path and (3) continuous rotational symmetry. Deformation on the cylinder surface is subject to rotation and will become slanted to the rotation axis. The fact (3) requires the deformation be periodic. According to fact (2) and elastic mechanics, the crease will be a straight line and appear at the place where deformation is the biggest\cite{Kresling}. Combined with fact (1), we are left with only two possibilities for the periodic crease pattern, either parallel or triangular. The former can be ruled by an interesting exercise: create a slanted array of parallel ridge/valley pairs on a A4 paper. You will find it impossible to roll it into a cylinder unless the creases are parallel to $L$ which contradicts the requirement that they ought to be slanted.

Geometry dictates that the triangular patten in Fig. \ref{alpha_theory} (a, b) obeys
\begin{equation}
	\begin{cases}
	\alpha+\beta+\gamma=\pi\\
	\alpha-\beta+\gamma=(1-\frac{2}{N})\pi
	\end{cases}
\label{beta}\end{equation}
which gives $\beta=\pi/N$. We expect $N$ comes from the competition of two potential energies.  First is the energy stored in ridges and valleys, which clearly favors a small $N$. According to the simulation results in Fig. \ref{material} (c), this energy is proportional to the crease length. To counter-balance it, we need to find a second source that prefers a large $N$. The aforementioned topological mechanism\cite{topology} rules out the interactions between ridges/valleys. We notice that, when the originally circular boundaries are deformed to polygons, the vicinity of each edge will be buckled. Presumably the longer the edge  $2\pi R/N$, the more severe this buckling. We expect the overlap of buckling from edges that face each other from opposite boundaries can be viewed as a repulsive potential. The strength of this interaction is assumed to be inversely proportional to their distance. Overall, the  energy consists of
\begin{equation}
	E(N)=N\epsilon\Big(\frac{L}{\sin\alpha}+\frac{L}{\sin\gamma}\Big)+N\eta\Big(\frac{2\pi R}{N}\Big)^2\Big(\frac{\sin\alpha}{L}\Big)
\label{energy}\end{equation}
where $\epsilon$ and $\eta$ are phenomenological parameters.

Use the law of sines and the empirical fact that $N\gg 1$, we can approximate
\begin{equation}
	\alpha\approx\sin^{-1}\sqrt{\frac{L}{2R}}
\label{alpha}\end{equation}
Detail proof of Eq. (\ref{alpha}) can be found in SM. By using Eqs. (\ref{beta},  \ref{alpha}), we can minimize Eq. (\ref{energy}) and obtain
\begin{equation}
	N\approx\pi\sqrt{\frac{\eta R}{\epsilon L}}
	\label{N}
\end{equation}
which now depends only on $\eta/\epsilon$. This expression describes well the trend in Fig. \ref{material} (a). Apparently the experimentally observed $N$ has to be an integer. Since the predicted value in Eq. (\ref{N}) is mostly not an integer, it implies an intrinsic randomness for $N$ in our system. For instance, $N=10.5$ can result in either $N=10$ or $11$.

Although experimentally $N$ is not sensitive to external tension $T$ and thickness $t$, their influence can be incorporated by adding a dimensionless correction $T/(Yt^2)$ inside the square root of Eq. (\ref{N}) where  $Y$ is Young's modulus. This is consistent with our experimental data in SM that $N$ is relatively more sensitive to $t$ than $Y$. 

The boundary of CO transition is at $R/L\approx 0.5$. This is obvious from Eq. (\ref{alpha}) because $\sin\alpha \leq 1$. In other words, it comes from the geometric constraint. An intuitive way to imagine this critical value is that the projection of the longer crease in Fig. \ref{alpha_theory}(b) on the base should never exceed twice the circumradius of the polygon in order for the structure to be fully foldable. 
\begin{equation}
\frac{L}{\sin\alpha}\leq 2R\frac{\pi/N}{\sin(\pi/N)}
\label{radius}
\end{equation}
Incorporating Eq. (\ref{N}) in Eq. (\ref{radius}) gives a more rigorous expression for the threshold:
\begin{equation}
	\frac{R}{L}\geq\frac{1}{2}(1+\frac{2\epsilon}{3\eta})
\end{equation}
In origami state, ridge-valley pairs are stable and arranged into predestined sites. In contrast, the excessive length of creases breaks the stable structure and renders their arrangement deviant from the triangles. This frustration of geometry then leads to local buckling. From this perspective of symmetry, the origami state can be viewed as an outcome of the reduction of continuous to discrete rotational symmetry. In the mean time, both rotational and twisting symmetries are broken by the emergence of local buckling in the crumple state.

{\sl Simulations.---} Besides Fig. \ref{material} (a, b), we search for more evidence of the CO transition from the energetics of creases and their interactions from molecular dynamics (MD) simulation. We construct the cylindrical shell by rolling up a thin sheet which was modelled by the triangular lattice with bond length $r=1$. The bending and stretching moduli are set to be comparable to realistic ones. The system is composed of 60150 beads in total. The Lennard-Jones potential is used to ensure that no bead can penetrate each other. During the simulation, we give every atom on both boundaries a constant angular velocity, making them rotate oppositely. We also let two boundaries move freely in the axial direction.  The simulation result for number $N$ versus $R/L$, plotted in Fig. \ref{material} (a), matches the experimental data. 

Another proof of the CO transition is provided in the ratio of bending to stretching energies $E_b/E_s$ for the whole cylinder, which changes from 1 to 3 in Fig. \ref{material} (c). Further evidence can also be found in the same figure, which shows  a dip in the transition for the exponent $b$ in the scaling relation $E(\ell)\sim \ell^b$ for the total energy  $E$ stored in crease of length $\ell$. To verify the validity of this relation, we build a graphical user interface that can show colorful energy-density plots and be used to integrate the total energy of beads on any line destined by the user. After sampling the creases, we use the least squares fitting  to estimate $b$ and its standard error. 

Our simulations do not find any bending energy in the facets between creases, which supports the topological mechanism surmised by Ref. \cite{topology} from 1-D. In addition to this, the simulation also demonstrates there is buckling in the vicinity of edges, as described by Eq. (\ref{energy}).

\begin{figure}[!h]
	\centering
	\includegraphics[width=8cm]{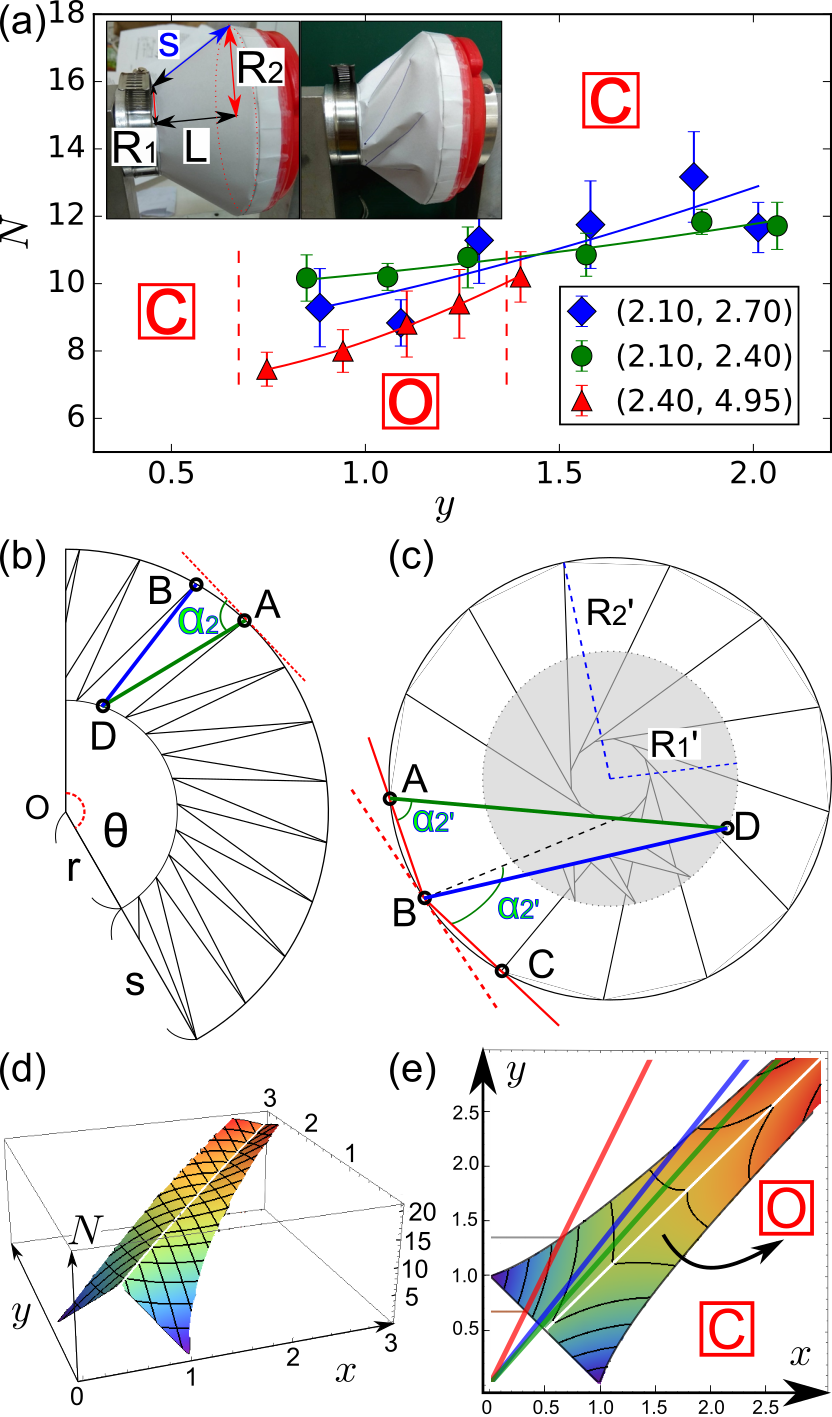}
	\caption
	{(color online) The boundary of double CO transition for $(R_1,R_2)=(2.40,4.95)$ is marked by red-dash line in panel (a) which is the projection of experimental data $N(x,y)$ onto $y$-axis. It also demonstrates that our theoretical curves consist to the experiment results. Schematic plots in (b, c) exhibit the geometry of creases after and before a twisted cone is unfolded. The 3-D surface in panel (d) describes the exact solution of $N(x,y)$. In panel (e), the red/blue/green line is correspond to $(R_1,R_2)=(2.40,4.95)/(2.10,2.70)/(2.10,2.40)$, while the black line depicts the state boundary $(1-x^2+y^2)^2\leq(x+y)(1+y-x)^2$.
	}\label{sample} 
\end{figure}

{\sl Generalization.---} Next, we verify whether twisting or continuous symmetry defined in the section of theoretical model triggers the CO transition. This can be tested by using more general Euclidean membranes, such as a truncated conical shell. Different from cylinders, a truncated cone has three parameters, $R_1, R_2$ and $s$, defined in inset of Fig. \ref{sample} (a). From dimensional analysis and previous experience, we expect the degree of freedom can be reduced to $x \equiv {R_1}/{s}$ and $y \equiv {R_2}/{s}$. As expected, we found the buckling to become irregular when $s$ is too long. What's surprising is that this irregular state reoccurs at short $s$. In other words, there exist two CO transitions, as marked by the dashed lines in Fig. \ref{sample} (c) for $(R_1,  R_2)=(2.40, 4.95)$.

To clarify the source of this second transition, we have to resort to theoretical analysis. It turns out that Eqs. (\ref{beta},  \ref{energy}, \ref{alpha}) are still valid except for some modifications as delineated in SM. The revised  version of Eq. (\ref{energy}) becomes 
\begin{equation}
	E(N)=N\epsilon({\ell_{v} + \ell_{r}})+N\eta\Big(\frac{2\pi R_1}{N}\Big)\Big(\frac{2\pi R_2}{N}\Big)\frac{\sin\alpha_2}{s}
\label{energy2}\end{equation}
where $\ell_{v}$/$\ell_{r}$ denote the length of valley/ridge and $\alpha_2$ is defined in Fig. \ref{sample} (b). After some algebra, we obtain the analytic solution:
\begin{equation}
N(x,y)\approx
\begin{dcases}
\pi\sqrt{\frac{x^2(x+y)(1+y-x)^2}{2y^2}\big(\frac{\eta}{\epsilon}\big)},\ y\ge x\\
\pi\sqrt{\frac{y^2(x+y)(1+x-y)^2}{2x^2}\big(\frac{\eta}{\epsilon}\big)},\ y\le x
\end{dcases}.
	\label{N-truncated}
\end{equation}
and plot Fig. \ref{sample}(d), which is consistent to experimental data in (a). Detailed derivations can be found in SM. 

An intuitive way to understand the second transition is to imagine valley $\overline{AD}$ and ridge $\overline{BD}$ that connect the two coaxial circles in Fig. \ref{sample} (c). Apparently, geometry requires them to obey
\begin{equation}
s\leq\overline{BD}\leq\overline{AD}\leq (R_1+R_2)\frac{\pi/N}{\sin(\pi/N)}
\end{equation}
which reduces to Eq. (\ref{radius}) when $R_1=R_2$. After some calculations, this inequality can be simplified to: 
\begin{equation}
(1-x^2+y^2)^2\leq(x+y)(1+y-x)^2
\end{equation}
which constrains the valid range of $x, y$ for origami state. Solid lines in Fig. \ref{sample} (e) mark the state boundaries. 

\begin{figure}[!h]
	\centering
	\includegraphics[width=8cm]{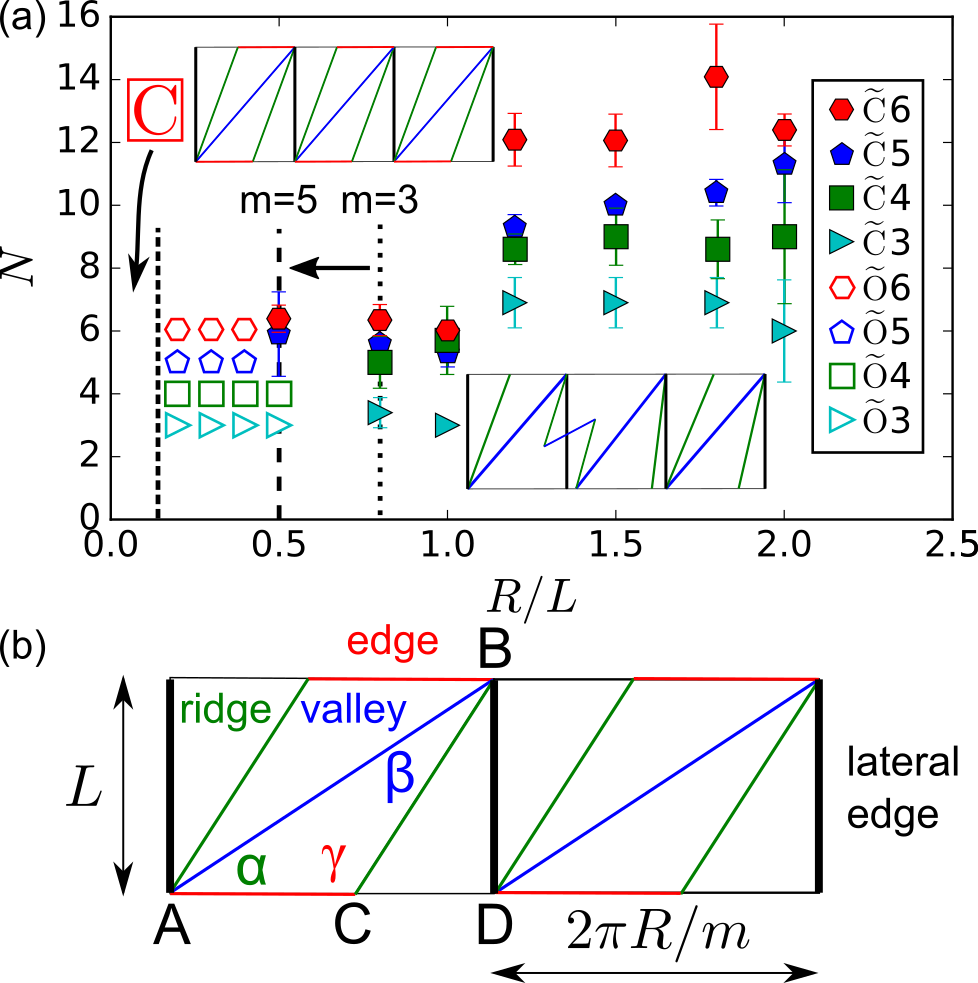}
	\caption
	{(color online) Panel (a) shows the experimental data of $m$-polygon where $m$ = 3, 4, 5, 6. The legends $\rm\widetilde{C}$/$\rm\widetilde{O}$ denote crumple-like/origami-like states. The black arrow indicates how the boundary of $\rm\widetilde{O}\widetilde{C}$ transition decreases as m increases. The upper/lower inset shows the schematic pattern of creases in $\rm\widetilde{O}$/$\rm\widetilde{C}$ states. Panel (b) is the magnified view of upper inset.
	}\label{polygon} 
\end{figure}

By now, we have shown that fact (1) in the section of theoretical model is not crucial for CO transition by using truncated cones. How about fact (3)? In the following we shall lessen the continuous rotational symmetry to a discrete one by replacing the circular bases by $m$-polygons with circumference $2\pi R$. The experimental results are shown in Fig. \ref{polygon} (a). When $R/L$ is small, the creases are periodic with $N=m$, as plotted schematically in the left inset. But the pattern becomes irregular as $R/L$ gets too large, as in the right inset. We label these two states as $\rm\widetilde{O}$ and $\rm\widetilde{C}$ which can be distinguished by whether errorbars $\delta N$ equal zero or not. Note that they are not only distinct from the CO states in Figs. (\ref{material}) and (\ref{sample}), but their order during the transition is reversed.

Before giving a physical picture for the occurrence of $\rm\widetilde{O}$ and $\rm\widetilde{C}$ states, we need to clarify the role of lateral edges, as $\overline{BD}$ in Fig. \ref{polygon} (b). They store bending energy with fold or dihedral angle $\phi_p=\pi-2\pi/m$ through localized plastic deformations\cite{response of crease}. The dislocation of lateral edge will exclude\cite{Landau} ridges and valleys from crossing it. This fact leads to $N=m$ in $\rm\widetilde{O}$ state. Based on the behavior of cylinders, $N$ tends to increase with $R/L$ in O state. We believe this dilemma between $\rm\widetilde{O}$ and O causes the non-periodic pattern in $\rm\widetilde{C}$ state that is dissimilar to C.
Likewise, experience on cylinders also informs us that a C state must await at small $R/L$. As a result, polygon cylinders can exhibit double CO transitions ($\rm C\widetilde{O}\widetilde{C}$).

Now we need to address the importance of $m$. A smaller $m$ gives smaller $\phi_p$ and stronger exclusive interaction with ridges/valleys due to larger dislocation and deformation. Consequently, the range of $\rm\widetilde{O}$ will shrink when $m$ increases, as shown in Fig. \ref{polygon} (a). It is thus expected that $\rm\widetilde{O}$ will disappear at some threshold $m^*$  beyond which the influence of lateral edges becomes negligible and  $\rm\widetilde{C}$ will be converted to O state. In other words, the polygon with $m\geq m^*$ behaves no different from cylinders and exhibits only one crumple-origami transition (CO).

{\sl Outlook.---} We presented the first example of coexistence between origami and crumple states and demonstrated the diversities of their transition in various Euclidean membranes. The distinctions between these two states are represented by the regularities of their creases, acoustic emissions, energetic ratios $E_b/E_s$, and exponents of $E\sim\ell^b$. Our theory deduces that the instability and randomness in crumple state come from frustration of geometry. The connection between our phenomenological parameters and traditional material properties remains to be explored. It will be fruitful to focus on $\rm C\widetilde{O}\widetilde{C}$ transition and the switch between $\rm C\widetilde{O}\widetilde{C}$ and CO, such as finding $m^*$, detail mechanism of pattern-formation, geometric conditions, and determining the phase diagram. In addition, the conjectured boundary of $\rm C\widetilde{O}$ needs to be verified in future experiments.  Compared to truncated cones that fix $L$ but not  $R$, it is reasonable to further investigate oblique circular cylinders that fix $R$ while varying $L$.  Generalizing this research to non-Euclidean membranes, such as spheres, will be another challenge in both theoretical analyses and experimental realization. Potential application may be to include the role of torsions on the orogeny, besides the conventional theory of compression\cite{orogeny-all, orogeny-comp} and volcano activities\cite{orogeny-all}.

We gratefully acknowledge financial support from MoST in Taiwan under grant 108-2112-M007-011-MY3.

\end{document}